\documentclass[pre,twocolumn,superscriptaddress,nofootinbib]{revtex4}

\usepackage{graphicx,chemarr,color}

\begin{document}

\title{Biochemical feedback and its application to immune cells II: dynamics and critical slowing down}

\author{Tommy A.\ Byrd}
\thanks{These authors contributed equally.}
\affiliation{Department of Physics and Astronomy, Purdue University, West Lafayette, Indiana 47907, USA}

\author{Amir Erez}
\thanks{These authors contributed equally.}
\affiliation{Department of Molecular Biology, Princeton University, Princeton, NJ 08544, USA}

\author{Robert M.\ Vogel}
\affiliation{IBM T.\ J.\ Watson Research Center, Yorktown Heights, New York 10598, USA}

\author{Curtis Peterson}
\affiliation{Department of Physics and Astronomy, Purdue University, West Lafayette, Indiana 47907, USA}
\affiliation{Department of Physics and School of Mathematical and Statistical Sciences, Arizona State University, Tempe, Arizona  85287}

\author{Michael Vennettilli}
\affiliation{Department of Physics and Astronomy, Purdue University, West Lafayette, Indiana 47907, USA}

\author{Gr\'egoire Altan-Bonnet}
\affiliation{Immunodynamics Group, Cancer and Inflammation Program, National Cancer Institute, National Institutes of Health, Bethesda, Maryland 20814, USA}

\author{Andrew Mugler}
\email{amugler@purdue.edu}
\affiliation{Department of Physics and Astronomy, Purdue University, West Lafayette, Indiana 47907, USA}

\begin{abstract}
Near a bifurcation point, the response time of a system is expected to diverge due to the phenomenon of critical slowing down. We investigate critical slowing down in well-mixed stochastic models of biochemical feedback by exploiting a mapping to the mean-field Ising universality class. This mapping allows us to quantify critical slowing down in experiments where we measure the response of T cells to drugs. Specifically, the addition of a drug is equivalent to a sudden quench in parameter space, and we find that quenches that take the cell closer to its critical point result in slower responses. We further demonstrate that our class of biochemical feedback models exhibits the Kibble-Zurek collapse for continuously driven systems, which predicts the scaling of hysteresis in cellular responses to more gradual perturbations. We discuss the implications of our results in terms of the tradeoff between a precise and a fast response.
\end{abstract}

\maketitle

\section{Introduction}

Critical slowing down is the phenomenon in which the relaxation time of a dynamical system diverges at a bifurcation point \cite{strogatz2018nonlinear}. Biological systems are inherently dynamic, and therefore one generally expects critical slowing down to accompany transitions between their dynamic regimes. Indeed, signatures of critical slowing down, including increased autocorrelation time and increased fluctuations, have been shown to precede an extinction transition in many biological populations \cite{scheffer2009early, scheffer2012anticipating}, including bacteria \cite{veraart2012recovery}, yeast \cite{dai2012generic}, and entire ecosystems \cite{wang2012flickering}. Similar signatures are also found in other biological time series, including dynamics of protein activity \cite{sha2003hysteresis} and neural spike dynamics \cite{meisel2015critical}.

Canonically, critical slowing down depends on scaling exponents that define divergences along particular parameter directions in the vicinity of a critical point \cite{hohenberg1977theory}. Therefore, connecting the theory of critical slowing down to biological data requires identification of thermodynamic state variables, their scaling exponents, and a principled definition of distance from the critical point. However, in most biological systems it is not obvious how to define the thermodynamic state variables, let alone scaling exponents and distance from criticality. In a previous study \cite{erez2018universality} we showed how near its bifurcation point, a class of biochemical systems can be mapped to the mean-field Ising model, thus defining the state variables and their associated scaling exponents. This provides a starting point for the investigation of critical slowing down in such systems, as well as how to apply such a theory to experimental data.

Additionally, most studies of critical slowing down in biological systems investigate the response to a sudden experimental perturbation (a ``quench''), such as a dilution or the addition of a nutrient or drug. This leaves unexplored the response to gradual environmental changes, a common natural scenario. When a gradual change drives a system near its critical point, critical slowing down delays the system's response such that no matter how gradual the change, the response lags behind the driving. In physical systems this effect is known as the Kibble-Zurek mechanism \cite{kibble1976topology, zurek1985cosmological}, which predicts these nonequilibrium lagging dynamics in terms of the exponents of the critical point. It remains unclear whether and how the Kibble-Zurek mechanism applies to biological systems.

Here we investigate critical slowing down for well-mixed biochemical networks with positive feedback, and we use our theory to interpret the response of immune cells to an inhibitory drug. Using our previously derived mapping \cite{erez2018universality}, we show theoretically that critical slowing down in our class of models proceeds according to the static and dynamic exponents of the mean-field Ising universality class. The mapping identifies an effective temperature and magnetic field in terms of the biochemical parameters, which defines a distance from the critical point that can be extracted from experimental fluorescence data. We find that drug-induced quenches that take an immune cell closer to its critical point result in longer response times, in qualitative agreement with our theory. We then show theoretically that our system, when driven across its bifurcation point, falls out of steady state in the manner predicted by the Kibble-Zurek mechanism, thereby extending Kibble-Zurek theory to a biologically relevant nonequilibrium setting. Our work elucidates the effects of critical slowing down in biological systems with feedback, and provides insights for interpreting cell responses near a dynamical transition point.

\section{Results}

\begin{figure}[t]
\centering
\includegraphics[width=\linewidth]{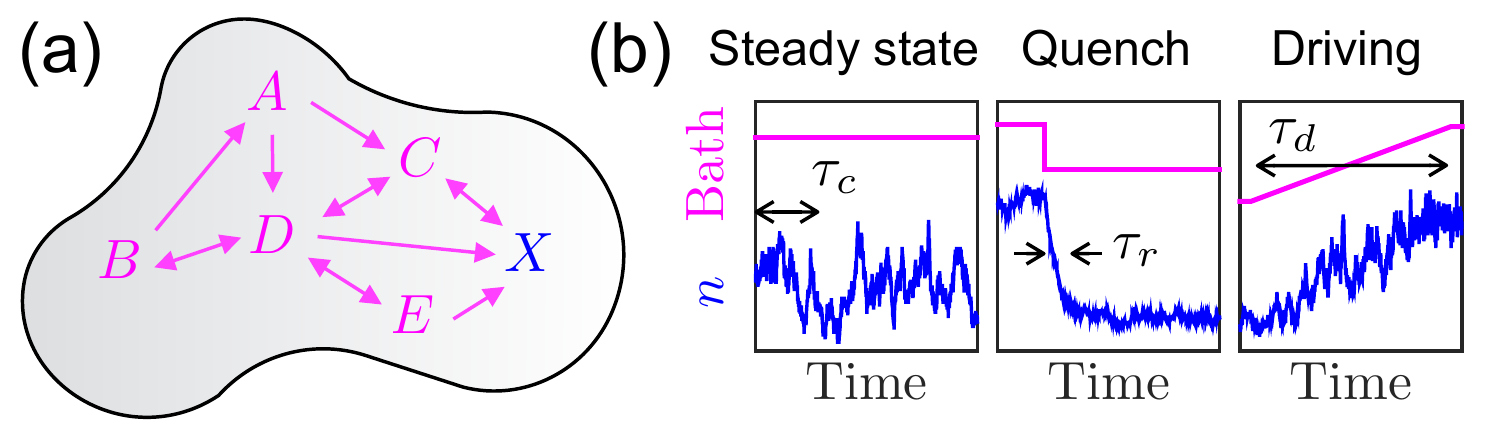}
\caption{(a) Inside a cell, a chemical species $X$ with molecule number $n$ exists in a bath of other species. (b) We consider steady-state, quench, and driven dynamics for the bath, and focus on the autocorrelation time $\tau_c$, response time $\tau_r$, and driving time $\tau_d$, respectively.}
\label{fig:setup}
\end{figure}

We consider a well-mixed reaction network in a cell where $X$ is the molecular species of interest, and the other species $A$, $B$, $C$, etc.\ form a chemical bath for $X$ [Fig.\ \ref{fig:setup}(a)]. Whereas previously we considered only the steady state distribution of $X$ \cite{erez2018universality}, here we focus on dynamics in and out of steady state. Specifically, as shown in Fig.\ \ref{fig:setup}(b), we consider (i) steady state, where the bath is constant in time; (ii) a quench, where the bath changes its parameters suddenly; and (iii) driving, where the bath changes its parameters slowly and continuously. In each case we are interested in a corresponding timescale: (i) the autocorrelation time $\tau_c$ of $X$, (ii) the response time $\tau_r$ of $X$, and (iii) the driving time $\tau_d$ of the bath.

First we review the key features of our stochastic framework for biochemical feedback and its mapping to the mean-field Ising model \cite{erez2018universality}. We consider an arbitrary number of reactions $r$ in which $X$ is produced from bath species $Y_r^\pm$ and/or $X$ itself (feedback),
\begin{equation}
\label{eq:rxns}
j_rX + Y_r^+ \rightleftharpoons (j_r+1)X + Y_r^-,
\end{equation}
where $j_r$ are stoichiometric integers. The probability of observing $n$ molecules of species $X$ in steady state according to Eq.\ \ref{eq:rxns} is
\begin{equation}
\label{eq:pn}
p_n = \frac{p_0}{n!} \prod_{j=1}^n f_j,
\end{equation}
where $p_0^{-1} = \sum_{n=0}^\infty(1/n!)\prod_{j=1}^n f_j$ is set by normalization, and $f_n$ is a nonlinear feedback function governed by the reaction network. The inverse of Eq.\ \ref{eq:pn},
\begin{equation}
\label{eq:fn}
f_n = \frac{np_n}{p_{n-1}},
\end{equation}
allows calculation of the feedback function from the distribution. The function $f_n$ determines an effective order parameter, reduced temperature, and magnetic field,
\begin{equation}
\label{eq:cparam}
m \equiv \frac{n_*-n_c}{n_c}, \quad
h \equiv \frac{2(f_{n_c} - n_c)}{-f'''_{n_c}n_c^3}, \quad
\theta \equiv \frac{2(1-f'_{n_c})}{-f'''_{n_c}n_c^2},
\end{equation}
respectively, where $n_c$ is defined by $f''_{n_c} = 0$, and $n_*$ are the maxima of $p_n$. Qualitatively, $n_c$ sets the typical molecule number, $\theta$ drives the system to a unimodal ($\theta > 0$) or bimodal ($\theta < 0$) state, and $h$ biases the system to high ($h > 0$) or low ($h < 0$) molecule numbers. The critical point occurs at $\theta = h = 0$. The state variables $m$, $\theta$, and $h$ scale according to the exponents $\alpha=0$, $\beta=1/2$, $\gamma=1$, and $\delta=3$ of the mean-field Ising universality class. Detailed analysis of this mapping in steady state is found in our previous work \cite{erez2018universality}.

Near the critical point, all specific realizations of a class of systems scale in the same way, and therefore it suffices to consider a particular realization of Eq.\ \ref{eq:rxns} from here on. We choose Schl\"ogl's second model \cite{erez2018universality}, a simple and well-studied case \cite{schlogl1972chemical, dewel1977renormalization, nicolis1980systematic, brachet1981critical, grassberger1982phase, prakash1997dynamics, liu2007quadratic, vellela2009stochastic} in which $X$ is either produced spontaneously from bath species $A$, or in a trimolecular reaction from two existing $X$ molecules and bath species $B$,
\begin{equation}
\label{eq:schlogl_rxns}
A \xrightleftharpoons[k_1^-]{k_1^+} X, \quad 2X+B \xrightleftharpoons[k_2^-]{k_2^+} 3X.
\end{equation}
In this case the feedback function is
\begin{equation}
\label{eq:schlogl_fn}
f_n = \frac{aK^2 + s(n-1)(n-2)}{(n-1)(n-2)+K^2},
\end{equation}
where we have introduced the dimensionless quantities $a \equiv k_1^+n_A/k_1^-$, $s \equiv k_2^+ n_B/k_2^-$, and $K^2 \equiv k_1^-/k_2^-$ in terms of the reaction rates and the numbers of $A$ and $B$ molecules. Given Eqs.\ \ref{eq:cparam} and \ref{eq:schlogl_fn}, the effective thermodynamic variables $n_c$, $\theta$, and $h$ can be written in terms of $a$, $s$, and $K$ or vice versa \cite{erez2018universality}, with $1/k_1^-$ setting the units of time.

\subsection{Critical slowing down in steady state}

In steady state, critical slowing down causes correlations to become long-lived near a dynamical transition point. Qualitatively, the fixed point is transitioning from stable to unstable, and therefore the basin of attraction is becoming increasingly wide. As a result, a dynamic trajectory takes increasingly long excursions from the mean, making it heavily autocorrelated. The autocorrelation time $\tau_c$ diverges at the critical point according to \cite{pathria2011statistical}
\begin{align}
\label{eq:tauc1}
\tau_c|_{h=0} &\sim |\theta|^{-\nu z}, \\
\label{eq:tauc2}
\tau_c|_{\theta=0} &\sim |h|^{-\nu z/\beta\delta},
\end{align}
where we expect $\nu z = 1$ for mean-field dynamics \cite{hohenberg1977theory, kopietz2010introduction}. Here the autocorrelation time $\tau_c$ is defined as
\begin{equation}
\label{eq:tauc_def}
\tau_c = \frac{1}{\kappa(0)}\int_0^\infty dt\ \kappa(t),
\end{equation}
where $\kappa(t) = \langle n(0)n(t)\rangle - \bar{n}^2$ is the steady-state autocorrelation function, $\kappa(0) = \sigma^2$ is the variance, and we have taken the start time to be $t=0$ without loss of generality because the system is in steady state.

\begin{figure}[t]
\centering
\includegraphics[width=\linewidth]{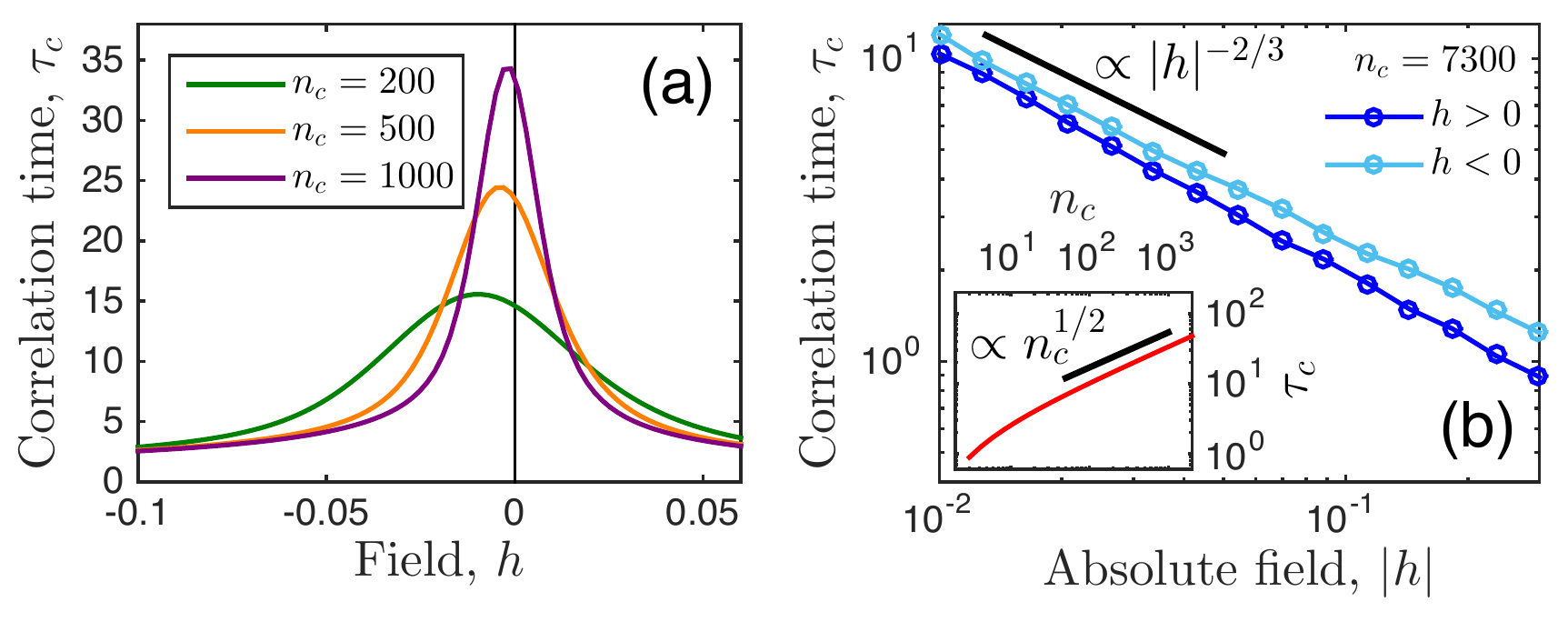}
\caption{Critical slowing down in steady state. (a) Autocorrelation time $\tau_c$ in Schl\"ogl model (Eq.\ \ref{eq:tauc_def}) peaks with field $h$ when reduced temperature $\theta = 0$. Height increases and location moves to $h=0$ as molecule number $n_c$ increases. Time is in units of $1/k_1^-$. (b) At large $n_c$, $\tau_c$ scales with $|h|$ with expected exponent of $\nu z/\beta\delta = 2/3$. Inset: $\tau_c$ at $\theta=h=0$ scales as $n_c^{1/2}$. In a and inset of b, $\tau_c$ is calculated using eigenfunctions with cutoff $N = \max(100,3n_c)$; in main panel of b, $\tau_c$ is calculated using batch means with 250 trajectories, duration $T = 10^5$, and batch time $\tau_b =$ 2,222 (see Appendix \ref{app:time}).}
\label{fig:ss}
\end{figure}

To confirm the value of $\nu z$, we plot $\tau_c$ vs.\ $h$ at $\theta = 0$ (Eq.\ \ref{eq:tauc2}). We calculate $\tau_c$ either directly from the master equation or from stochastic simulations \cite{gillespie1977exact} using the method of batch means \cite{thompson2010comparison} (see Appendix \ref{app:time}). The results are shown in Fig.\ \ref{fig:ss}. We see in Fig.\ \ref{fig:ss}(a) that $\tau_c$ indeed diverges with $h$, and that the location of the divergence approaches the expected value $h = 0$ as the molecule number $n_c$ increases. We also see that the height of the peak increases with $n_c$ due to the rounding of the divergence \cite{stephens2013statistical}. The inset of Fig.\ \ref{fig:ss}(b) plots this dependence: we see that $\tau_c$ at the critical point $\theta = h = 0$ scales like $n_c^{1/2}$ for large $n_c$ (the application of this dependence to dynamic driving will be discussed in Section \ref{sec:KZ}). Finally, we see in the main panel of Fig.\ \ref{fig:ss}(b) that when $n_c$ is sufficiently large, $\tau_c$ falls off with $|h|$ with the expected scaling exponent of $\nu z/\beta\delta = 2/3$. Taken together, these results confirm that the divergence of the autocorrelation time in the Schl\"ogl model obeys the static exponents of the mean-field Ising universality class ($\beta\delta = 3/2$) and the dynamic expectation for mean-field systems ($\nu z = 1$).

\subsection{Quench response and application to immune cells}

\begin{figure}
\centering
\includegraphics[width=\linewidth]{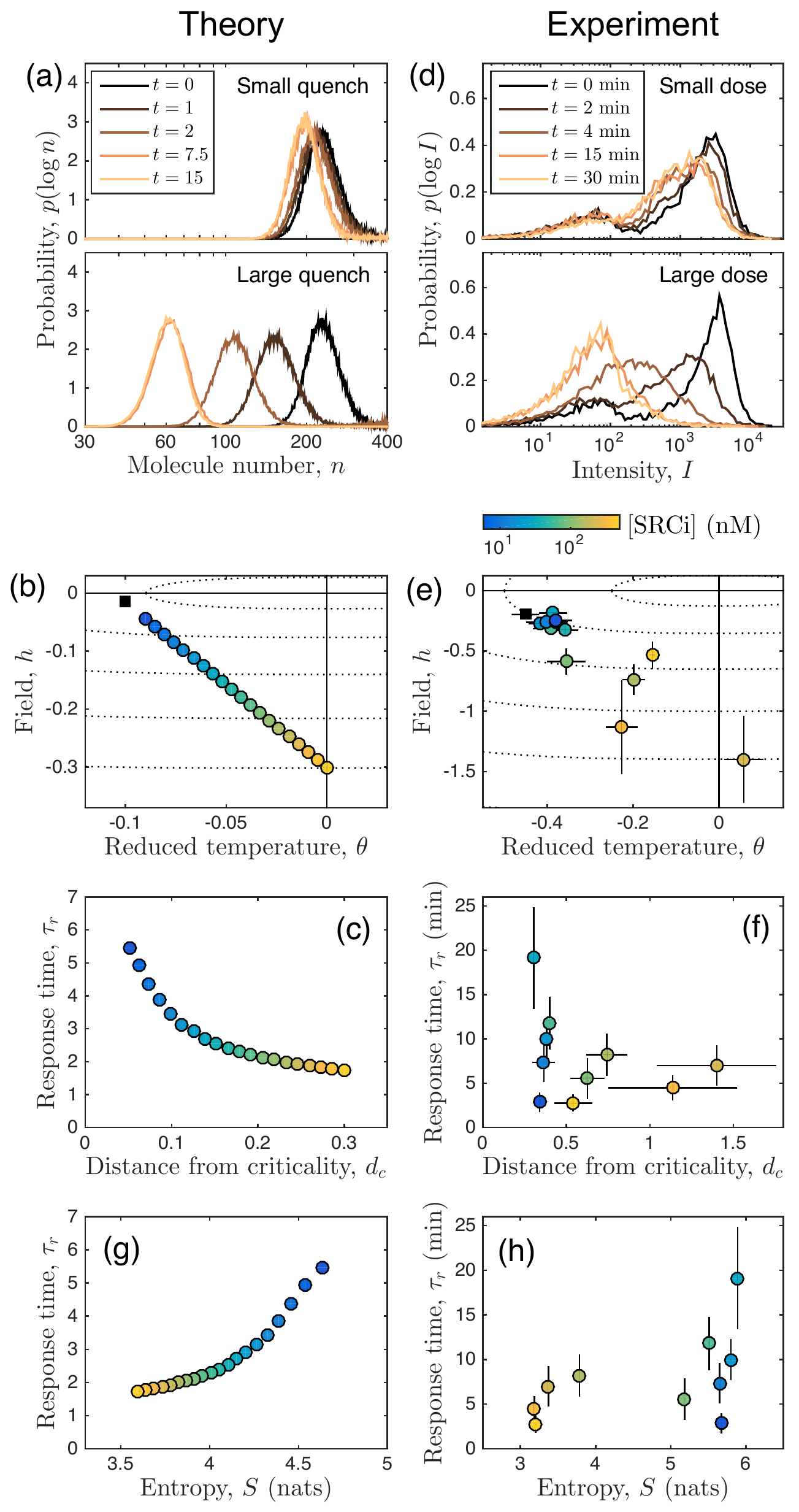}
\caption{Quench response in theory (left) and in immune cell experimental data (right). (a) Stochastic simulations of Shl\"ogl model show effect of small and large parameter quenches on distribution. Time is in units of $1/k_1^-$. (b) Initial (black square) and quenched (colored circles) parameter values in $\theta$ and $h$ space in model; $n_c = 500$. Dotted lines show contours of equal $d_c$ (Eq.\ \ref{eq:dc}), distance from critical point ($\theta = h = 0$). Response time $\tau_r$ in model (c) decreases with $d_c$ and (g) increases with entropy $S$. (d) Experimental distributions of T cell ppERK fluorescence intensity measured at times after addition of SRC inhibitor (see Fig.\ \ref{fig:doses} for all doses). (e) $\theta$ and $h$ extracted from initial distribution (black square) and final distributions (colored circles) for all [SRCi] doses (color bar). Experimental response time $\tau_r$ (f) decreases with $d_c$ and (h) increases with $S$. Error bars: for $\theta$ and $h$, standard error from Savitzky-Golay \cite{savitzky1964smoothing} filter windows $25 \le W \le 35$ \cite{erez2018universality}; for $d_c$, propagated in quadrature from e; for $\tau_r$, standard deviation of Riemann sums spanning left- to right-endpoint methods to approximate integral in Eq.\ \ref{eq:taur2}. In h, fluorescence of one molecule set to $I_1 = 10$.}
\label{fig:quench}
\end{figure}

When subjected to a sudden environmental change (a quench), the system will take some finite amount of time to respond [Fig.\ \ref{fig:setup}(b), middle]. We expect that if a quench takes the system closer to its critical point, the response time should be longer due to critical slowing down. To make this expectation quantitative, we define the response time $\tau_r$ in terms of the dynamics of the mean molecule number $\bar{n}$ as
\begin{equation}
\label{eq:taur1}
\tau_r = \frac{1}{\Delta\bar{n}(0)}\int_0^{t_{\max}} dt\ \Delta\bar{n}(t),
\end{equation}
where the quench occurs at $t=0$, we define $\Delta \bar{n}(t) = \bar{n}(t) - \bar{n}(t_{\max})$, and we ensure that $t_{\max} \gg \tau_r$. We compute $\bar{n}(t)$ from the time-dependent distribution $p_n(t)$ using the stochastic simulations. Examples of $p_n(t)$ for a small and a large quench are shown in Fig.\ \ref{fig:quench}(a).

We define the distance from the critical point in terms of the state variables $\theta$ and $h$. Specifically, $\tau_c$ scales identically with $\theta^{\beta\delta}$ as it does with $h$ (Eqs.\ \ref{eq:tauc1} and \ref{eq:tauc2}), which defines the Euclidean distance $d_c$ from the critical point as
\begin{equation}
\label{eq:dc}
d_c = \left[(\theta^{\beta\delta})^2 + h^2\right]^{1/2}.
\end{equation}
This measure will be important when comparing with the experiments because, as opposed to in most condensed matter experiments, it is difficult in the biological experiments we describe to manipulate only one parameter ($\theta$ or $h$) independently of the other.

To test whether the response time increases with proximity to the critical point, we must define initial values $\theta_0$ and $h_0$ for the environment before the quench, and a series of values $\theta$ and $h$ for the environment after the quench that are varying distances from the critical point $\theta = h = 0$. There are many such choices for these values, but anticipating the experimental results that we will describe shortly, we choose the initial point (black square) and final points (colored circles) shown in Fig.\ \ref{fig:quench}(b). Dotted curves of equal $d_c$ are also shown, which make clear that larger quenches (yellow circles) take the system farther from the critical point than smaller quenches (blue circles). The dependence of $\tau_r$ on $d_c$ is shown in Fig.\ \ref{fig:quench}(c), and we see that indeed $\tau_r$ decreases as $d_c$ increases, or equivalently the response time increases with proximity to the critical point.

We now compare our theory with data from immune cells. We focus on the abundance in T cells of doubly phosphorylated ERK (ppERK), a protein that initiates cell proliferation and is implicated in the self/non-self decision between mounting an immune response or not \cite{vogel2016dichotomy, altan2005modeling}. Specifically, we use flow cytometry to measure the ppERK distribution at various times after the addition of a drug that inhibits SRC, a key enzyme in the cascade that leads to ERK phosphorlyation (see Appendix \ref{app:expt} for experimental methods). When the dose of the drug is small, the distribution hardly changes [Fig.\ \ref{fig:quench}(d), top]; whereas when the dose is large, the distribution changes significantly [Fig.\ \ref{fig:quench}(d), bottom]. The responses to all doses are shown in Appendix \ref{app:expt}.

After the addition of the drug, the cells reach a new steady-state ppERK distribution [green curves in Fig.\ \ref{fig:quench}(d)]. The distribution corresponds to an effective feedback function via Eq.\ \ref{eq:fn}, from which the effective temperature $\theta$ and field $h$ can be calculated via Eq.\ \ref{eq:cparam} \cite{erez2018universality}. The values of $\theta$ and $h$ calculated from the experimental distributions are shown in Fig.\ \ref{fig:quench}(e). We see that larger doses take the cells farther from their initial distribution (black square), as expected. We also see that larger doses take the system farther from the critical point $\theta = h = 0$. The general shape of the $\theta$ and $h$ values motivated our choice of theoretical values in Fig.\ \ref{fig:quench}(b).

We define the response time to the drug as in Eq.\ \ref{eq:taur1}, here in terms of the mean fluorescence intensity of ppERK,
\begin{equation}
\label{eq:taur2}
\tau_r = \frac{1}{\Delta\bar{I}(0)}\int_0^{t_{\max}} dt\ \Delta\bar{I}(t),
\end{equation}
where $\Delta \bar{I}(t) = \bar{I}(t) - \bar{I}(t_{\max})$ and $t_{\max} = 30$ min. We calculate the distance from criticality using Eq.\ \ref{eq:dc} as before, here using the experimental values of $\theta$ and $h$. We see in Fig.\ \ref{fig:quench}(f) that the response time $\tau_r$ decreases with the distance from criticality $d_c$, consistent with the prediction from the theory [Fig.\ \ref{fig:quench}(c)]. This suggests that critical slowing down occurs in the response of the T cells to the drug.

Although in Fig.\ \ref{fig:quench}(f) the response time $\tau_r$ comes directly from the experimental data, the distance from criticality $d_c$ is calculated from the experimental data using expressions from the theory (Eqs.\ \ref{eq:fn} and \ref{eq:cparam}). This makes the results in Figs.\ \ref{fig:quench}(c) and \ref{fig:quench}(f) not entirely independent. To confirm that the agreement between Figs.\ \ref{fig:quench}(c) and \ref{fig:quench}(f) is not a result of an implicit co-dependence, we seek a measure that is related to distance from criticality but that is not dependent on the theory. We choose the entropy of the distribution $S = -\sum_n p_n\log p_n$ because near criticality, the distribution is broad and flat, and therefore we expect the entropy to be large; whereas far from criticality, the distribution has either one or two narrow peaks, and therefore we expect the entropy to be small \cite{erez2018universality}. Indeed, we see in Fig.\ \ref{fig:quench}(g) that in the theory, the response time $\tau_r$ increases with the entropy $S$, consistent with the fact that it decreases with the distance from criticality [Fig.\ \ref{fig:quench}(c)]. The same is evident in the experiments: we see in Fig.\ \ref{fig:quench}(h) that low drug doses correspond to long response times and high entropies, whereas high drug doses correspond to short response times and low entropies, resulting in an increase of response time $\tau_r$ with entropy $S$. Calculating the entropy in Fig.\ \ref{fig:quench}(h) requires a conversion between intensity $I$ and molecule number $n$, and we have checked that the results in Fig.\ \ref{fig:quench}(h) are qualitatively unchanged for different choices of this conversion factor over several orders of magnitude. The agreement between Figs.\ \ref{fig:quench}(g) and \ref{fig:quench}(h) offers further evidence that the T cells experience critical slowing down, with the data analysis completely independent from our theory.

\subsection{Dynamic driving and Kibble-Zurek collapse}
\label{sec:KZ}

While some environmental changes are sudden, many changes in a biological context are gradual [Fig.\ \ref{fig:setup}(b), right]. When a gradual change drives a system through its critical point, critical slowing down delays the system's response such that no matter how gradual the change, the response lags behind the driving. Although in a biological setting the driving protocol could take many forms, terms beyond the leading-order linear term do not change the critical dynamics \cite{chandran2012kibble}. This is a major theoretical advantage because it allows us to specialize to linear driving without loss of biological realism. Specifically, we focus on linear driving across the critical point with driving time $\tau_d$, setting either $\theta(t) = \theta_i-(\theta_f - \theta_i)t/\tau_d$ and $h=0$, or $h(t) = h_i-(h_f-h_i)t/\tau_d$ and $\theta = 0$, where $i$ and $f$ denote the initial and final parameter values, respectively.

In a traditional equilibrium setting, the dynamics of lagging trajectories are described in terms of the critical exponents by the Kibble-Zurek mechanism \cite{kibble1976topology, zurek1985cosmological}. The idea of the Kibble-Zurek mechanism is that far from the critical point, the change in the system's correlation time due to the driving, over a correlation time, is small compared to the correlation time itself, $(d\tau_c/dt)\tau_c \ll \tau_c$, and therefore the system responds adiabatically. However, as the system is driven closer to the critical point, these two quantities are on the same order, or $d\tau_c/dt \sim 1$, and the system begins to lag. Applying this condition to Eqs.\ \ref{eq:tauc1} and \ref{eq:tauc2}, and using the above expressions for $\theta(t)$ and $h(t)$, one obtains
\begin{align}
\label{eq:kz1}
\theta &\sim \tau_d^{-1/(\nu z+1)}, \\
\label{eq:kz2}
h &\sim \tau_d^{-\beta\delta/(\nu z + \beta\delta)},
\end{align}
respectively. Because $m \sim (-\theta)^\beta$ or $m \sim h^{1/\delta}$ near criticality in the mean-field Ising class, we have \begin{align}
\label{eq:kzm1}
m &\sim \tau_d^{-\beta/(\nu z+1)}, \\
\label{eq:kzm2}
m &\sim \tau_d^{-\beta/(\nu z + \beta\delta)},
\end{align}
respectively. Therefore, if the system is driven at different timescales $\tau_d$, the Kibble-Zurek mechanism predicts that plots of the rescaled variables $m\tau_d^{\beta/(\nu z+1)}$ vs.\ $\theta\tau_d^{1/(\nu z+1)}$ or $m\tau_d^{\beta/(\nu z + \beta\delta)}$ vs.\ $h\tau_d^{\beta\delta/(\nu z + \beta\delta)}$ will collapse onto universal curves.

\begin{figure}[t]
\centering
\includegraphics[width=\linewidth]{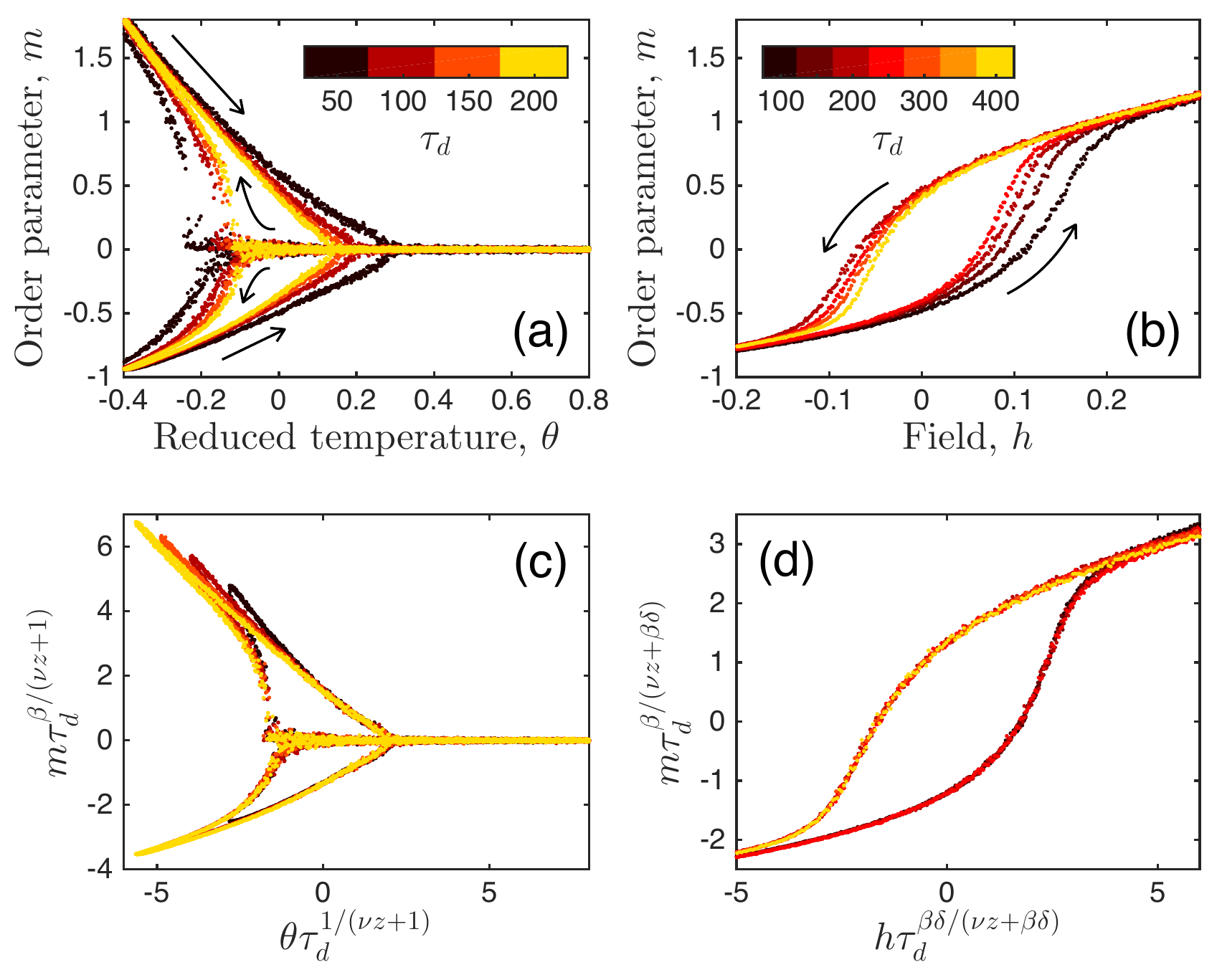}
\caption{Dynamic driving and Kibble-Zurek collapse. (a) As reduced temperature $\theta$ is driven over time $\tau_d$ in Schl\"ogl model, order parameter $m$ lags behind due to critical slowing down. Decreasing $\theta$ causes supercooling (left curves), while increasing $\theta$ causes superheating (right curves), resulting in hysteresis. (b) Same, for driving $h$. (c, d) Rescaled curves collapse as predicted. Each point is computed via Eq.\ \ref{eq:cparam} from the mode $n_*$ in b, or the modes $n_*^{(1)}<n_c$ and $n_*^{(2)}>n_c$ in a, of $10^5$ simulation trajectories. For finite-size correction we use $n_c = 10\tau_d$ in a and $n_c = 22\tau_d^{4/5}$ in b. Time is in units of $1/k_1^-$.}
\label{fig:kz}
\end{figure}

When testing these predictions using simulations of a spatially extended physical system, the finite size of the system causes a truncation of the autocorrelation time. This truncation is usually accounted for using a finite-size correction \cite{chandran2012kibble}. In our system, a similar truncation of the autocorrelation time is caused by the finite number of molecules. Specifically, the inset of Fig.\ \ref{fig:ss}(b) shows that at criticality we have $\tau_c \sim n_c^{1/2}$ for large $n_c$, where $n_c$ sets the typical number of molecules in the system. Therefore, we interpret $n_c$ as a ``system size,'' and we correct for finite-size effects in the following way. Combining the relation $\tau_c \sim n_c^{1/2}$ with Eqs.\ \ref{eq:tauc1} and \ref{eq:tauc2}, and Eqs.\ \ref{eq:kz1} and \ref{eq:kz2}, we obtain
\begin{align}
\label{eq:finite1}
n_c &\sim \tau_d^{2\nu z/(\nu z+1)},\\
\label{eq:finite2}
n_c &\sim \tau_d^{2\nu z/(\nu z + \beta\delta)},
\end{align}
for the driving of $\theta$ or $h$, respectively. We choose $n_c$ arbitrarily for a particular driving time $\tau_d$, and when we choose a new $\tau_d$, we scale $n_c$ appropriately according to Eqs.\ \ref{eq:finite1} and \ref{eq:finite2}.

This procedure allows us to test the predictions of the Kibble-Zurek mechanism using simulations of the Schl\"ogl model. The results are shown in Fig.\ \ref{fig:kz}. We see in Fig.\ \ref{fig:kz}(a) that as $\theta$ is driven from a positive to a negative value, the bifurcation response is lagging, occurring at a value less than the critical value $\theta = 0$ (supercooling). Conversely, when $\theta$ is driven from a negative to a positive value, the convergence occurs at a value greater than $\theta = 0$ (superheating). In both directions, the lag is larger when the driving is faster, corresponding to smaller values of $\tau_d$ (from yellow to dark brown). We see in Fig.\ \ref{fig:kz}(b) that similar effects occur for the driving of $h$. Yet, we see in Figs.\ \ref{fig:kz}(c) and (d) that the rescaled variables collapse onto single, direction-dependent curves within large regions near criticality. Note that the direction dependence (i.e., hysteresis) is preserved as part of these universal curves, but the lags vanish in the collapse. This result demonstrates that our nonequilibrium birth-death model exhibits the Kibble-Zurek collapse predicted for critical systems. Together with our previous findings, this result suggests that such a collapse should emerge in biological experiments where environmental parameters (e.g., drug dose) are dynamically controlled in a gradual manner. More broadly, by phenomenologically collapsing such experimental curves, it should be possible to deduce the critical exponents of such biological systems without fine-tuning them to criticality, but instead by gradual parameter
sweeps.

\section{Discussion}

We have investigated critical slowing down in a minimal stochastic model of biochemical feedback.  By exploiting a mapping to Ising-like thermodynamic variables, we have made quantitative predictions for the response of a system with feedback to both sudden and gradual environmental changes. In response to a sudden change (a quench), we have shown that the system will respond more slowly if the quench takes it closer to its critical point, in qualitative agreement with multiple-time-point flow cytometry experiments in immune cells. In response to more gradual driving, we have shown that the lagging dynamics of the system proceed according to the Kibble-Zurek mechanism for driven critical phenomena. Together, our results elucidate the consequences of critical slowing down for biochemical systems with feedback, and demonstrate those consequences on an example system from immunology.

For the immune cells, critical slowing down may present a tradeoff in terms of the speed vs.\ the precision of an immune response. Specifically, ppERK is implicated in the decision of whether or not to mount the immune response \cite{vogel2016dichotomy, altan2005modeling}, suggesting that ppERK dynamics near the bifurcation point are of key biological importance. Yet, the bifurcation point is the point where critical slowing down is most pronounced. In fact, the inset of Fig.\ \ref{fig:ss}(b) demonstrates that the system slows down as the number of molecules in the system increases. On the other hand, large molecule number is known to decrease intrinsic noise and thereby increase the precision of a response \cite{elowitz2002stochastic}. This suggests that cells may face a tradeoff in terms of speed vs.\ precision when responding to changes that occur near criticality, as suggested for other biological systems \cite{skoge2011dynamics, mora2011biological}.

Our work extends the Kibble-Zurek mechanism to a nonequilibrium biological context. Traditionally, the mechanism has been applied to physical systems from cosmology \cite{kibble1976topology} and from hard \cite{zurek1985cosmological, del2014universality} or soft \cite{deutschlander2015kibble} condensed matter. Here, we extend the mechanism to the context of biochemical networks with feedback, where the system already exists in a nonequilibrium steady state, and the external protocol takes the system further out of equilibrium into a driven state. It will be interesting to see to what other nonequilibrium contexts the Kibble-Zurek mechanism can be successfully applied \cite{deffner2017kibble}.

The theory we present here assumes only intrinsic birth-death reactions and neglects more complex mechanisms such as bursting \cite{friedman2006linking, mugler2009spectral}, parameter fluctuations \cite{shahrezaei2008colored, horsthemke1984noise}, or cell-to-cell variability \cite{cotari2013cell, erez2018modeling} that may play an important role in the immune cells. Nonetheless, similar models that also focus only on intrinsic noise have successfully described ppERK in T cells in the past \cite{das2009digital, prill2015noise}. Moreover, we expect that intrinsic fluctuations should play their largest role near the bifurcation point. Finally, we expect that near the bifurcation point, the essential behavior of the system should be captured by any model that falls within the appropriate universality class.

In this and previous work \cite{erez2018universality} we have explored the dynamic and static scaling properties of single cells subject to biochemical feedback. Natural extensions include generalizing the theory to cell populations or other systems that are not well-mixed such as intracellular compartments. This would allow one to investigate the spatial consequences of proximity to a bifurcation point, such as long-range correlations in molecule numbers and the associated implications for sensing, information transmission, patterning, or other biological functions.


\section*{Acknowledgments}
We thank Anushya Chandran for helpful communications. This work was supported by Simons Foundation grant 376198 (T.A.B.\ and A.M.), Human Frontier Science Program grant LT000123/2014 (Amir Erez), National Science Foundation Research Experiences for Undergraduates grant PHY-1460899 (C.P.), National Institutes of Health (NIH) grants R01 GM082938 (A.E.) and R01 AI083408 (A.E., R.V., and G.A.-B) and the NIH National Cancer Institute Intramural Research programs of the Center for Cancer Research (A.E.\ and G.A.-B.).

\appendix

\section{Autocorrelation time}
\label{app:time}

We calculate the autocorrelation time $\tau_c$ (Eq.\ \ref{eq:tauc_def}) for the Schl\"ogl model in steady state using one of two methods, the first more efficient for small molecule numbers, and the second more efficient for large molecule numbers. The first method is to calculate $\tau_c$ numerically from the master equation for $p_n$ by eigenfunction expansion. The master equation follows from the reactions in Eq.\ \ref{eq:schlogl_rxns} \cite{erez2018universality} and can be written in vector notation as
\begin{equation}
\label{eq:meL}
\dot{\vec{p}} = {\bf L}\vec{p}.
\end{equation}
where ${\bf L}$ is a tridiagonal matrix containing the birth and death propensities for $X$. The eigenvectors of {\bf L} satisfy
\begin{align}
{\bf L}\vec{v}_j &= \lambda_j \vec{v}_j, \\
\vec{u}_j{\bf L} &= \lambda_j \vec{u}_j,
\end{align}
where the eigenvalues obey $\lambda_j \le 0$ with only $\lambda_0$ vanishing for the steady state, and $\vec{v}_j^T \neq \vec{u}_j$ because {\bf L} is not Hermitian \cite{walczak2009stochastic}. Because Eq.\ \ref{eq:meL} is linear in $\vec{p}$, the solution is
\begin{equation}
\label{eq:pt}
p_n(t) = \sum_{jn'} u_{jn'} p_{n'}(0) e^{\lambda_jt} v_{nj}
\end{equation}
for initial condition $p_n(0)$. Calling $n(0) \equiv m$ and $n(t) \equiv n$, we write the autocorrelation function (see Eq.\ \ref{eq:tauc_def}) as
\begin{equation}
\label{eq:kappa2}
\kappa(t) = -\bar{n}^2 + \sum_{mn} p_{mn} mn = -\bar{n}^2 + \sum_{mn} p_{n|m}p_m mn,
\end{equation}
where $p_m = v_{m0}$ is the steady-state distribution, and $p_{n|m}$ is the dynamic solution at time $t$ assuming the system starts with $m$ molecules. That is, $p_{n|m}$ is given by Eq.\ \ref{eq:pt} with initial condition $p_n(0) = \delta_{nm}$. Eq.\ \ref{eq:kappa2} becomes
\begin{align}
\kappa(t) &= -\bar{n}^2 + \sum_{mn} mn v_{m0} \sum_j u_{jm} e^{\lambda_jt} v_{nj} \\
\label{eq:kappa3}
	&= \sum_{mn} mn v_{m0} \sum_{j=1}^\infty u_{jm} e^{\lambda_jt} v_{nj},
\end{align}
where the second step uses orthonormality, $\sum_j v_{nj} u_{jn'} = \delta_{nn'}$, and probability conservation, $u_{0n} = 1$, to recognize that the $j=0$ term evaluates to $\bar{n}^2$. Inserting Eq.\ \ref{eq:kappa3} into Eq.\ \ref{eq:tauc_def} and performing the integral (recalling that $\lambda_j < 0$ for $j > 0$), we obtain
\begin{equation}
\tau_c = \frac{1}{\sigma^2} \sum_{mn} mn v_{m0}
	\sum_{j=1}^\infty u_{jm} \left(\frac{1}{-\lambda_j}\right) v_{nj}.
\end{equation}
\begin{figure}
\centering
\includegraphics[width=\linewidth]{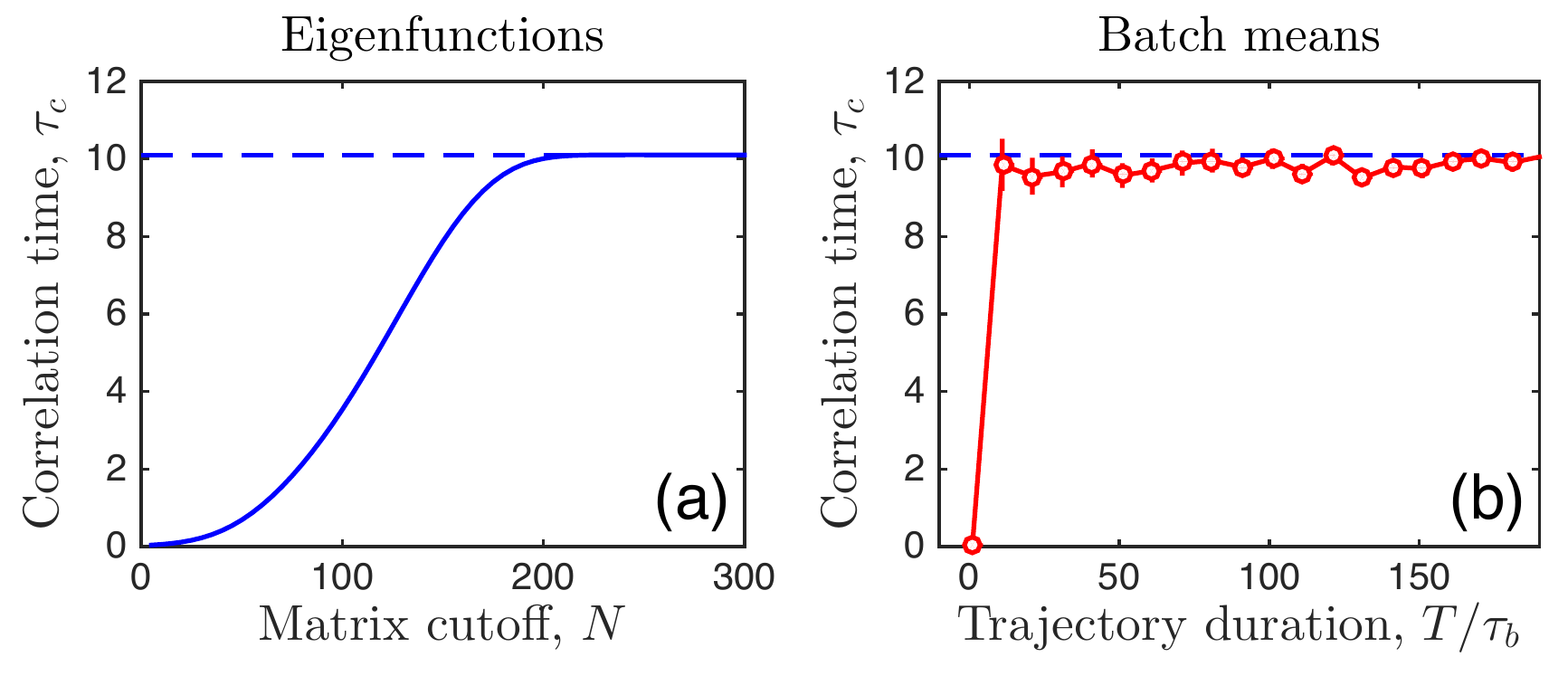}
\caption{Autocorrelation time computed (a) numerically using eigenfunction expansion or (b) by simulation using method of batch means. For sufficient cutoff $N$ or trajectory duration $T$, respectively, both methods converge to same value (dashed line). Parameters: $\theta = h = 0$ and $n_c = 100$. Time is in units of $1/k_1^-$. In (b), $\tau_b = 1000$, and error bars are standard error from $50$ trajectories.}
\label{fig:tauc}
\end{figure}
In matrix notation,
\begin{equation}
\label{eq:tauc_mat}
\tau_c = \sigma^{-2} \vec{n} {\bf V} {\bf F} {\bf U} \vec{w},
\end{equation}
where $\vec{n}$ is a row vector, $\vec{w} = mv_{m0}$ is a column vector, and neither the eigenvector matrices ${\bf V}$ and ${\bf U}$ nor the diagonal matrix $F_{jj'} = -\delta_{jj'}/\lambda_j$ contain the $j=0$ term. Numerically, we compute $\tau_c$ via Eq.\ \ref{eq:tauc_mat} using a cutoff $N > n_c$ for the vectors and matrices.

The second method is to calculate $\tau_c$ from stochastic simulations \cite{gillespie1977exact} and the method of batch means \cite{thompson2010comparison}. The idea is to divide a simulation trajectory of length $T$ into batches of length $\tau_b$. In the limit $T\gg\tau_b\gg\tau_c$, the correlation time can be estimated by \cite{thompson2010comparison}
\begin{equation}
\label{eq:batch}
\tau_c = \frac{\tau_b\sigma_b^2}{2\sigma^2},
\end{equation}
where $\sigma_b^2$ is the variance of the means of the batches.

In Fig.\ \ref{fig:tauc} we verify that the two methods converge to the same limit for sufficiently large $N$ or $T$, respectively. We find that the first method is more efficient until $n_c \sim 1000$, when numerically computing the eigenvectors for large $N > n_c$ becomes intractable.

\section{Experimental methods}
\label{app:expt}

\begin{figure*}[t]
\centering
\includegraphics[width=.7\linewidth]{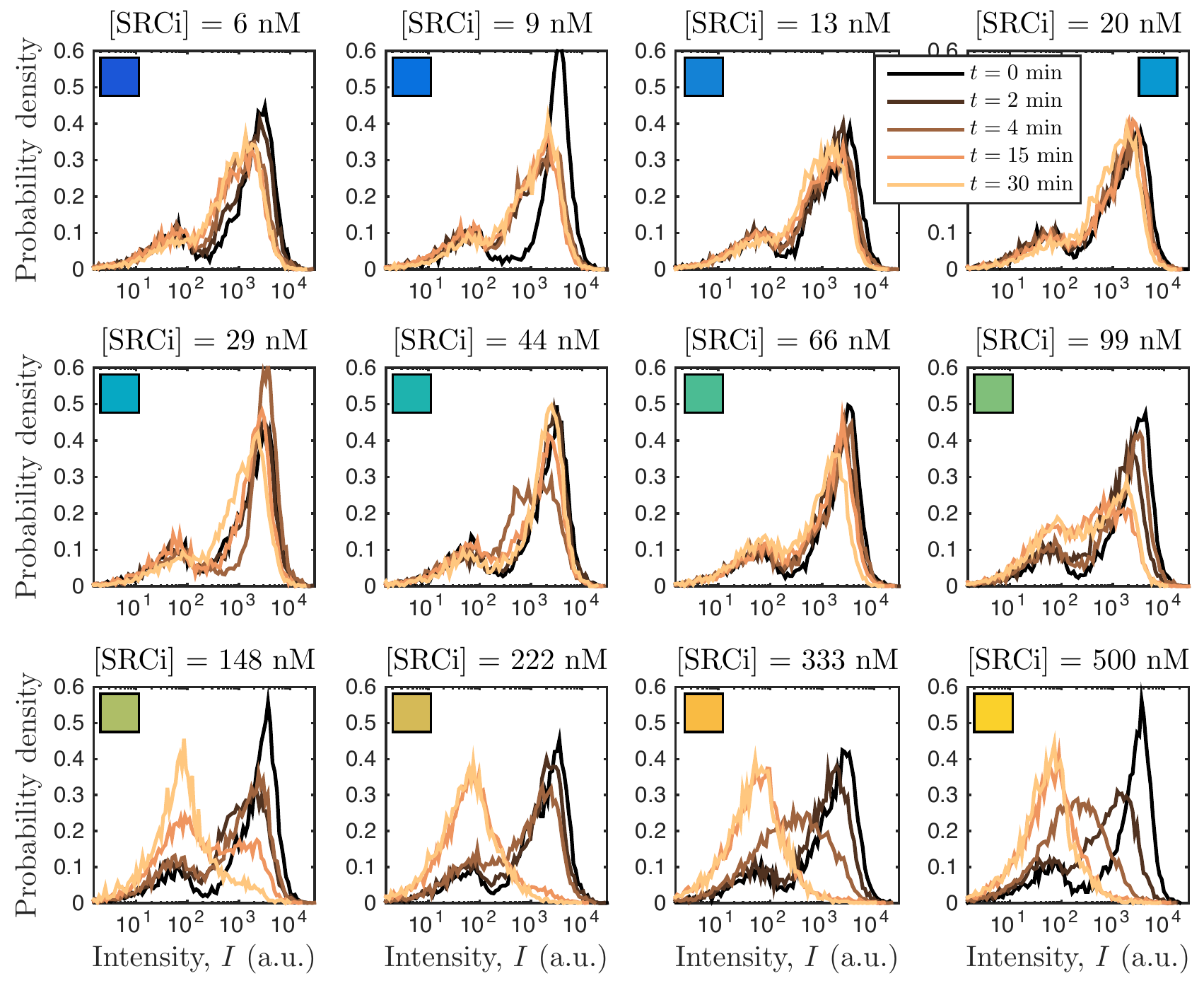}
\caption{Experimental distributions of T cell ppERK fluorescence intensity measured at times after addition of SRC inhibitor. Times given in legend in upper right. Dose given in title of each panel; colored square in upper corner of each panel corresponds to color bar in Fig.\ \ref{fig:quench}(e), (f) and (h). Panels with smallest and largest dose are reproduced in Fig.\ \ref{fig:quench}(d).}
\label{fig:doses}
\end{figure*}

The data in Fig.\ \ref{fig:doses} [of which the smallest and largest doses are reproduced in Fig.\ \ref{fig:quench}(d)] were acquired at the same time and in a similar way as the data published in \cite{vogel2016dichotomy} and summarized in \cite{erez2018universality}. The difference is that, instead of only recording the data after steady state was reached, the time series was sampled by applying a chemical fixative to stop chemical reactions and preserve all biomolecular states. Specifically, we administered ice cold formaldehyde in PBS to each experimental well of a 96 well-v-bottom plate such that the final working dilution is 2\%, and then transferred the cell-fixative solution to a new 96 well-v-bottom plate on ice. Cells were kept on ice for 10 minutes and then precipitated by centrifugation, resuspended in ice-cold 90\% methanol, and placed in a $-20$ $^{\rm o}$C freezer until measurements were taken.


\end{document}